The following is the entire tex file
which should be compiled with the standard latex macro.
There is no figure.
----------------------------------------------------------------------
\documentstyle[12pt]{article}
\newcommand{\bra}{\langle}
\newcommand{\ket}{\rangle}
\newcommand{\di}{\mbox{d}}
\newcommand{\e}{\mbox{e}}
\newcommand{\k}{\mbox{i}}
\newcommand{\sk}{\mbox{\scriptsize i}}
\newcommand{\bm}[1]{\mbox{\boldmath{$ #1 $}}}
\makeatletter
\@addtoreset{equation}{section}
\makeatother
\renewcommand{\theequation}{\thesection.\arabic{equation}}
\begin{document}
%
%%%%%%%%%%%%%%%%%%%% title %%%%%%%%%%%%%%%%%%%%%%% sect 0 %%%%%%%%%%%%
%
\begin{flushright}
	DPNU-93-09
	\\
	March 1993
\end{flushright}
\vspace{12mm}
\begin{center}
	{\Large\bf
	Gauge Field, Parity and Uncertainty Relation
	\\
	\vspace{6mm}
	of Quantum Mechanics on $ S^1 $
	}
	\footnote{To be published in {\it Progress of Theoretical Physics}.}
	\\
	\vspace{12mm}
	{\large Shogo Tanimura}
	\footnote{e-mail address : tanimura@eken.phys.nagoya-u.ac.jp}
	\\
	\vspace{6mm}
	{\large \it Department of Physics, Nagoya University, }
	\\
	\vspace{4mm}
	{\large \it Nagoya 464-01, Japan}
	\\
\vspace{24mm}
{\bf Abstract}
\\
\vspace{6mm}
\begin{minipage}[t]{130mm}
	\baselineskip 6mm
	We consider the uncertainty relation
	between position and momentum of a particle on $ S^1 $
	(a circle).
	Since $ S^1 $ is compact,
	the uncertainty of position must be bounded.
	Consideration on the uncertainty of position demands
	delicate treatment.
 	Recently Ohnuki and Kitakado have formulated
	quantum mechanics on $ S^D $ (a $D$-dimensional sphere).
	Armed with their formulation, we examine this subject.
	We also consider parity and
	find a phenomenon similar to the spontaneous symmetry breaking.
	We discuss problems which we encounter
	when we attempt to formulate quantum mechanics on a general manifold.
\end{minipage}
\end{center}
%
%%%%%%%%%%%%%%%%%%%%%%%%%%%%%%%%%%%%%%%%%%%%%%%%%% sect 1 %%%%%%%%%%%%
%
\newpage
\baselineskip 7mm
\section{Introduction}
We consider the uncertainty relation between position and momentum
of a particle on $ S^1 $, that is a circle.
Since $ S^1 $ is compact, the uncertainty of the position must be bounded.
However, how should we define the uncertainty of the position?
Can we use the angle variable $ \theta $ to indicate the position?
Can we define the uncertainty by the variance of the angle variable,
$ \bra \Delta \theta^2 \ket = \bra \theta^2 \ket - \bra \theta \ket^2 $,
as usual?
We must carefully use the angle variable because of its multi-valuedness ;
$ \theta + 2 \pi n $ ($ n $ : integer)
indicates the same point as the one which $ \theta $ does.
Many authors \cite{Judge}, \cite{Bouten}, \cite{Carruthers}
have been discussing this subject.
It is also known \cite{Carruthers} that the angle variable cannot become
a well-defined quantum-mechanical operator.
\par
Recently Ohnuki and Kitakado \cite{Ohnuki}
formulated quantum mechanics (QM) on $ S^D $,
that is a $ D $-dimensional sphere.
They defined the fundamental algebra of operators
without reliance on angle variables
and they have classified irreducible representations of the algebra.
They showed that
{\it there are an infinite number of inequivalent representations}
in contrast to QM on an Euclidean space.
For QM on an Euclidean space, we regard the canonical commutation relation
$ [\: x_j , p_k \:] = \k \delta_{jk} $ as the fundamental algebra.
It is known as von Neumann's theorem \cite{Neumann} that
the irreducible representation of the canonical commutation relation
is {\it uniquely} determined up to a unitary equivalent class.
In other words, various QM can exist on a sphere
while only single one exists on an Euclidean space.
\par
Armed with their formulation,
we would like to approach the uncertainty relation on a sphere.
In this paper we consider only the case of $ S^1 $.
We shall begin with a review of their formulation
and proceed to discuss a gauge field and a path integral on $ S^1 $.
Next we shall examine an energy spectrum of a free particle on $ S^1 $
and its symmetry.
We shall identify this symmetry with parity.
Next we shall examine the uncertainty relation on $ S^1 $
and define a minimum uncertainty wave packet.
Furthermore, we examine a behavior of the minimum uncertainty wave packet
when the radius of $ S^1 $ increases to infinity.
Finally, we shall discuss problems which we will encounter
when we attempt to formulate QM on a general manifold.
%
%%%%%%%%%%%%%%%%%%%%%%%%%%%%%%%%%%%%%%%%%%%%%%%%%% sect 2 %%%%%%%%%%%%
%
\section{Quantum Mechanics on $ S^1 $}
\subsection{Definition and construction}
First, we review QM on $ S^1 $ formulated by Ohnuki and Kitakado.
Its definition is as follows :
\begin{enumerate}
	\renewcommand{\labelenumi}{(\roman{enumi})}
	\item
		$ G $ is a self-adjoint operator
		and $ W $ is a unitary operator.
	\item
		These operators satisfy the commutation relation
		\begin{equation}
			[ \: G , W \: ] = W .
			\label{eqn:2.1}
		\end{equation}
\end{enumerate}
$ G , W $ and $ W^{\dagger} $ generate an algebra,
which is denoted by $ \cal A $.
They called $ \cal A $ the fundamental algebra for QM on $ S^1 $.
To construct QM one needs not only an algebra but also its representation.
In other words, QM is defined as a set of observables (operators)
and state vectors (operands = representation space
= Hilbert space $ \cal H $).\footnote{In the strict sense,
I regard a set $( {\cal A}, {\cal H}, H)$ as QM, where $ H $ is a Hamiltonian.}
\par
They constructed representations as follows.
Since $ G $ is a self-adjoint operator,
thus it has an eigenvector with a real eigenvalue $ \alpha $ :
\begin{equation}
	G \, | \alpha \ket = \alpha \, | \alpha \ket ,
	\qquad
	\bra \alpha | \alpha \ket = 1 .
	\label{eqn:2.2}
\end{equation}
$ W $ raises the eigenvalues of $ G $ :
\begin{eqnarray}
	G \, W | \alpha \ket
	&=&
	( \, [ \: G , W \: ] + W G ) \, | \alpha \ket
	\nonumber\\
	&=&
	( \, W + W \alpha ) \, | \alpha \ket
	\nonumber\\
	&=&
	( 1 + \alpha ) W | \alpha \ket .
	\label{eqn:2.3}
\end{eqnarray}
Inversely, $ W^{-1} = W^{\dagger} $ lowers them. We define
\begin{equation}
	| n + \alpha \ket := W^n \, | \alpha \ket ,
	\quad
	( n = 0 , \pm 1 , \pm 2 , \cdots )
	\label{eqn:2.4}
\end{equation}
which have the following properties :
\begin{eqnarray}
	&&
	G \, | n + \alpha \ket = ( n + \alpha ) \, | n + \alpha \ket ,
	\label{eqn:2.5}
	\\
	&&
	\bra m + \alpha | n + \alpha \ket = \delta_{mn} .
	\label{eqn:2.6}
\end{eqnarray}
The latter follows from self-adjointness of $ G $ and unitarity of $ W $.
With a fixed real number $ \alpha $, we define a Hilbert space
$ {\cal H}_{\alpha} $ by completing the vector space of linear combinations of
$ | n + \alpha \ket $ ($ n $ : integer). Equation (\ref{eqn:2.5}) and
\begin{equation}
	W \, | n + \alpha \ket = | n + 1 + \alpha \ket
	\label{eqn:2.7}
\end{equation}
define an irreducible representation of $ \cal A $ on $ {\cal H}_{\alpha} $.
\par
Here we state two propositions :
\begin{enumerate}
	\renewcommand{\labelenumi}{(\roman{enumi})}
	\item
		$ {\cal H}_{\alpha} $ and $ {\cal H}_{\beta} $
		are unitary equivalent representation spaces
		if and only if
		the difference between $ \alpha $ and $ \beta $ is an integer.
	\item
		Assume one has an arbitrary irreducible representation space
		$ \cal H $ of $ \cal A $,
		then there exists a real number $ \alpha $
		such that $ \cal H $ is unitary equivalent to
		$ {\cal H}_{\alpha} $.
\end{enumerate}
Therefore the classification of irreducible representations of $ \cal A $
has been completed;
the whole of inequivalent irreducible representation spaces is
$ \{ {\cal H}_{\alpha} \} \: ( 0 \leq \alpha < 1 ) $.
\subsection{Wave function}
In the previous section we have diagonalized $ G $.
We can also diagonalize $ W $.
So doing, we will understand that
$ ( {\cal A} , {\cal H}_{\alpha}) $ is indeed QM on $ S^1 $ .
\par
In what follows, we fix the representation space $ {\cal H}_{\alpha} $.
We shall consider the eigenvalue problem
\begin{equation}
	W \, | \xi \ket = \xi \, | \xi \ket .
	\label{eqn:2.8}
\end{equation}
Its solution is
\begin{eqnarray}
	&&
	| \theta \ket :=
	\kappa(\theta)
	\sum_{n=-\infty}^{\infty} \e^{-\sk n \theta} | n + \alpha \ket ,
	\label{eqn:2.9}
	\\
	&&
	W \, | \theta \ket = \e^{\sk \theta} | \theta \ket ,
	\label{eqn:2.10}
\end{eqnarray}
where $ \theta $ is a real number and $ \kappa(\theta) $ is a complex-valued
function such that
$ | \kappa(\theta) | = 1 , \: \kappa(\theta+2\pi) = \kappa(\theta) $.
One can easily verify the following:
\begin{enumerate}
	\renewcommand{\labelenumi}{(\roman{enumi})}
	\item
		periodicity
		\begin{equation}
			| \theta + 2 \pi \ket = | \theta \ket ,
			\label{eqn:2.11}
		\end{equation}
	\item
		orthonormality
		\begin{equation}
			\bra \theta | \theta' \ket
			= 2 \pi \sum_{n=-\infty}^{\infty}
			\delta( \theta - \theta' + 2 \pi n) ,
			\label{eqn:2.12}
		\end{equation}
	\item
		completeness
		\begin{equation}
			\int_0^{2 \pi} \frac{\di \theta}{2 \pi} \,
			| \theta \ket \bra \theta |
			= \sum_{n=-\infty}^{\infty}
			| n + \alpha \ket \bra n + \alpha |
			= 1_{\alpha}
			\quad ( \mbox{identity operator on }
			{\cal H}_{\alpha} ) ,
			\label{eqn:2.13}
		\end{equation}
	\item
		translation
		\begin{equation}
			\exp(-\k \lambda G) \, | \theta \ket
			= \e^{-\sk \lambda \alpha} \, \kappa(\theta) \,
			\kappa^*(\theta + \lambda) \, | \theta + \lambda \ket .
			\label{eqn:2.14}
		\end{equation}
\end{enumerate}
On account of these properties,
$ ( {\cal A}, {\cal H}_{\alpha} ) $ is indeed QM on $S^1$.
It is also reasonable to identify $ G $, and $ W $ as the momentum operator,
and the position operator, respectively.
\par
Let $ | \psi \ket $ be a state vector.
We define a wave function $ \psi(\theta) $ on $ S^1 $ by
\begin{equation}
	\psi(\theta) := \bra \theta | \psi \ket .
	\label{eqn:2.15}
\end{equation}
Inner product of Eq. (\ref{eqn:2.14}) with $ | \psi \ket $ is
\begin{equation}
	\bra \theta | \exp(\k \lambda G) | \psi \ket
	= \e^{\sk \lambda \alpha} \, \kappa^*(\theta) \,
	\kappa(\theta + \lambda) \, \bra \theta + \lambda | \psi \ket .
	\label{eqn:2.16}
\end{equation}
Differentiating both sides with respect to $ \lambda $
and putting $ \lambda=0 $, we obtain
$$
	\k \, \bra \theta | \, G \, | \psi \ket
	= \k \alpha \, \psi (\theta)
	+ \kappa^*(\theta) \, \frac{\partial \kappa(\theta)}{\partial \theta}
	\, \psi(\theta)
	+ \frac{\partial}{\partial\theta} \, \psi(\theta) ,
$$
that is
\begin{equation}
	\bra \theta | \, G \, | \psi \ket
	=
	\left[
	-\k \frac{\partial}{\partial\theta}
	- \k \kappa^*(\theta) \,
	\frac{\partial \kappa(\theta)}{\partial \theta}
	+ \alpha
	\right]
	\, \psi (\theta) .
	\label{eqn:2.17}
\end{equation}
On the other hand, from Eq. (\ref{eqn:2.10}) we obtain
\begin{equation}
	\bra \theta | \, W \, | \psi \ket
	= \e^{\sk\theta} \, \psi(\theta) .
	\label{eqn:2.18}
\end{equation}
The inner product $ \bra \chi | \psi \ket $ is expressed
in terms of wave functions as
\begin{equation}
	\bra \chi | \psi \ket =
	\int_0^{2 \pi} \frac{\di \theta}{2 \pi} \,
	\chi^*(\theta) \, \psi(\theta).
	\label{eqn:2.19}
\end{equation}
Equations (\ref{eqn:2.17}) and (\ref{eqn:2.18}) define a representation of
$ \cal A $ on the Hilbert space
$ L_2 (S^1) $, that is the space of square integrable functions on $ S^1 $.
\subsection{Gauge field}
We can now appreciate the physical meaning of the parameter $ \alpha $.
We define
\begin{equation}
	A(\theta) :=
	-\k \kappa^*(\theta) \,
	\frac{\partial \kappa(\theta)}{\partial \theta}
	+ \alpha ,
	\label{eqn:2.20}
\end{equation}
and call it a gauge field.
Soon we will see that this naming is reasonable.
In Eq. (\ref{eqn:2.9}), $ \kappa(\theta) $ remains arbitrary.
We can change it to $ \kappa'(\theta) $, which is related to
$ \kappa(\theta) $ by
\begin{equation}
	\kappa(\theta) = \omega(\theta) \, \kappa'(\theta) ,
	\label{eqn:2.21}
\end{equation}
where $ | \omega(\theta) | = 1, \: \omega(\theta + 2\pi) = \omega(\theta) $.
Thus the position eigenstates are transformed as
\begin{equation}
	| \theta \ket = \omega(\theta) \, | \theta \ket' ,
	\label{eqn:2.22}
\end{equation}
and the transformed wave function
$ \psi'(\theta) := \: '\bra \theta | \psi \ket $ is
\begin{equation}
	\psi'(\theta) = \omega(\theta) \, \psi(\theta).
	\label{eqn:2.23}
\end{equation}
Hence the operator $ G $ acts on $ \psi'(\theta) $ as
\begin{eqnarray}
	&&
	'\bra \theta | \, G \, | \psi \ket
	=
	\left[
	-\k \frac{\partial}{\partial\theta} + A'(\theta)
	\right]
	\, \psi' (\theta) ,
	\label{eqn:2.24}
	\\
	&&
	A'(\theta) = A(\theta) +
	\k \omega^*(\theta) \,
	\frac{\partial \omega(\theta)}{\partial \theta}.
	\label{eqn:2.25}
\end{eqnarray}
The pair of Eqs. (\ref{eqn:2.23}) and (\ref{eqn:2.25})
is nothing but a gauge transformation.
\par
Here we prove two propositions :
\begin{enumerate}
	\renewcommand{\labelenumi}{(\roman{enumi})}
	\item
		Assume that $ A(\theta) $ is an arbitrary real-valued function
		such that
		$ A(\theta + 2 \pi) = A(\theta) $.
		Then there exists a gauge transformation which takes
		$ A(\theta) $ to a constant function
		$ A'(\theta) \equiv \alpha $.
	\item
		For two constant functions,
		$ A' \equiv \alpha $ and $ A'' \equiv \beta $,
		there exists a unique gauge transformation which connects them,
		if and only if $ (\beta - \alpha) $ is an integer.
\end{enumerate}
\par
Proof of (i) :
We put
\begin{eqnarray}
	&&
	\alpha =
	\int_0^{2 \pi} \frac{\di \theta}{2 \pi} \, A(\theta) ,
	\label{eqn:2.26}
	\\
	&&
	\omega(\theta) =
	\exp
	\left[
	-\k \int_0^{\theta} \di \varphi \,
	( \alpha - A(\varphi) )
	\right] .
	\label{eqn:2.27}
\end{eqnarray}
Thus the transformation (\ref{eqn:2.25}) gives $ A'(\theta) \equiv \alpha $.
\\
\par
Proof of (ii) :
Assume the existence of such a transformation $ \omega' $. It must satisfy
\begin{equation}
	\k \omega'^* \, \frac{\partial \omega'}{\partial \theta}
	= \beta - \alpha .
	\label{eqn:2.28}
\end{equation}
This equation has the solution
\begin{equation}
	\omega'(\theta) = \exp \left[ -\k ( \beta - \alpha ) \theta \right].
	\label{eqn:2.29}
\end{equation}
However, to make $ \omega' $ a periodic function,
$ ( \beta - \alpha ) $ must be an integer.
\par
Consequently, the whole of gauge-inequivalent---physically inequivalent
gauge fields is
$ \{ A_\alpha \equiv \alpha \} \: ( 0 \leq \alpha < 1 ) $.
In one-dimensional space, there is no magnetic field
because the magnetic field is an antisymmetric tensor $ F_{ij} $,
which vanishes identically in one-dimension.
Nevertheless, the gauge field can exist.
It produces observable effects on a quantum system as is seen below.
\par
For instance, we will consider a free particle on $ S^1 $.
The eigenvalue problem of the Hamiltonian
\begin{equation}
	H = \frac 12 \, G^{\: 2}
	\label{eqn:2.30}
\end{equation}
is trivial. The solution is
\begin{equation}
	H \, | n + \alpha \ket = \frac 12 ( n + \alpha )^2 | n + \alpha \ket .
	\label{eqn:2.31}
\end{equation}
Apparently, the spectrum depends on the parameter $ \alpha $.
For $ \alpha = m $ ($ m $ : integer),
all the eigenvalues but one of the ground state are doubly degenerate.
While for $ \alpha = m + \frac 12 $,
the all eigenvalues are doubly degenerate.
For other values of $ \alpha $, there is no degeneracy.
It seems that these degeneracies reflect a {\it symmetry};
the Hamiltonian is invariant under the parity transformation $ G \to -G $.
We will discuss this point in the next section.
As $ n+\alpha = (n-1)+(\alpha+1) $,
the spectrum of $ H $ on $ {\cal H}_\alpha $ is same as that on
$ {\cal H}_{\alpha + 1} $.
Moreover, as $ (n+\alpha)^2 = (-n-\alpha )^2 $,
the spectrum of $ H $ on $ {\cal H}_\alpha $
is same to that on $ {\cal H}_{- \alpha} $, too.
Therefore the distinguishable values of $ \alpha $
range over $ 0 \le \alpha \le \frac 12 $.
By measuring the energy levels, an inhabitant of $ S^1 $ (Circulish!?)
can tell the value of $ \alpha $
with ambiguities which differ by integers and the sign.
Imagine that an inhabitant of a three-dimensional space
where $ S^1 $ is embedded looks at this tiny world $ S^1 $
and that he makes the magnetic field penetrating the hollow
surrounded by $ S^1 $.
Assume that the magnetic field is zero on $ S^1 $
and that the magnetic flux is $ 2 \pi \alpha $.
If the dear inhabitant of $ S^1 $ knew only classical mechanics,
he or she could not notice what has happened.
However, if he or she knew quantum mechanics and were clever enough,
he or she would perceive it.
\par
The gauge field influences not only the energy levels
but also probability amplitudes.
It has been known \cite{Ohnuki2} that if the Hamiltonian is of the form
\begin{equation}
	H = \frac 12 \, G^{\: 2} + V ( W, W^\dagger ) ,
	\label{eqn:2.32}
\end{equation}
the transition amplitude $ K $ is expressed in terms of path integral as
\begin{eqnarray}
	K(\theta' , \theta ; t)
	&:= &
	\bra \theta' | \exp( -\k H t ) | \theta \ket
	\nonumber
	\\
	& = &
	\sum_{n = -\infty}^{\infty}
	\int_{\mbox{\scriptsize winding} \: n \: \mbox{\scriptsize times}}
	{\cal D} \theta
	\, \exp( \k S_{\mbox{\scriptsize eff}} ) ,
	\label{eqn:2.33}
\end{eqnarray}
where the effective action is defined as
\begin{equation}
	S_{\mbox{\scriptsize eff}} := \int \di t
	\left[
	\frac 12 \biggl( \frac{\di \theta}{\di t} \biggr)^2
	- V ( \theta ) - \alpha \frac{\di \theta}{\di t}
	\right] .
	\label{eqn:2.34}
\end{equation}
In (\ref{eqn:2.33}),
integration over paths winding $ n $ times around $ S^1 $ and
summation with respect to the winding number are performed.
We would like to emphasize that the above path integral expression is derived
from the operator formalism.
It should be noticed that this global property---winding number---appears
from the operator formalism alone.
\par
The last term in (\ref{eqn:2.34}), $ \int \alpha \, \di \theta $ has no effect
on the equation of motion but has an observable effect on the amplitude.
To see the role of this term we rewrite (\ref{eqn:2.33}) as
\begin{equation}
	K(\theta' , \theta ; t) =
	\e^{ -\sk \alpha (\theta' - \theta) }
	\sum_{n = -\infty}^{\infty}
	\e^{ -\sk \alpha 2 \pi n }
	\int_{\mbox{\scriptsize winding} \: n \: \mbox{\scriptsize times}}
	{\cal D} \theta
	\, \exp( \k S_0 ),
	\label{eqn:2.33a}
\end{equation}
where $ S_0 = \int \di t \, [ \frac 12 \dot{\theta}^2 - V(\theta) ] $.
An amplitude for a path winding $ n $ times is weighted by
the phase factor $ \omega_n = \exp( -\k \alpha 2 \pi n ) $.
This phase factor causes observable interference effect;
this phenomenon is analogous to the Aharonov-Bohm effect.
Furthermore, $ \omega_n $'s obey composition rule;
$ \omega_m \, \omega_n = \omega_{m+n} $,
which says that $ m $-times winding followed by $ n $-times one is equal
to $ (m+n) $-times one.
According to \cite{Schulman}, \cite{Laidlaw},
$ \omega_n $ can be interpreted as a unitary representation
of the first homotopy group $ \pi^1(S^1) $.
\par
In $ S^1 $ there is physical electric field but no magnetic field.
Nonetheless, there exist quantum effects of the gauge field as is seen above.
This situation is in sharp contrast to the cases of dimensions higher than one.
\par
How about QM on $ S^D (D \ge 2) $?
Here we quote only a part of the results of Ohnuki and Kitakado
\cite{Ohnuki}, \cite{Kitakado}.
They set up the fundamental algebra for QM on $ S^D $
and constructed irreducible representations completely.
They showed that the spin degrees of freedom are naturally built in
and that the gauge field associated with the Wigner rotation
is inevitably introduced.
When $ D \ge 3 $, the group of gauge transformations is non-Abelian.
Furthermore, they found
that when $ D \ge 2 $, the field strength corresponding to the gauge field
does not vanish but has a monopole-like structure.
We recommend an interested reader to consult their article.
\subsection{Parity}
Let us now turn to the parity transformation on $ S^1 $.
Geometrically, the parity is defined as a operation inverting $ S^1 $
around a diameter.
If we measure an angular coordinate $ \theta $ from the axis fixed under
the parity operation, parity moves the point specified by $ \theta $
to the one specified by $ -\theta $.
In another expression, $ \e^{\sk \theta} \to \e^{-\sk \theta} $.
It is apparent that doubled parity is the identity operation,
that changes nothing.
Taking account of the above consideration, we postulate that
the parity operator $ P $ of QM on $ S^1 $ satisfies the following :
\begin{eqnarray}
	\mbox{(i)}
	&&
	P^{\dagger} P = 1 ,
	\label{eqn:2.35}
	\\
	\mbox{(ii)}
	&&
	P^2 = 1 ,
	\label{eqn:2.36}
	\\
	\mbox{(iii)}
	&&
	P^{\dagger} W P = W^{\dagger} .
	\label{eqn:2.37}
\end{eqnarray}
\par
What about the momentum operator?\footnote{Our style of description here
refers to J.J. Sakurai's book \cite{Sakurai}, p.252.}
The momentum $ G $ is like $ \di \theta / \di t $, so it is natural to expect
that it transforms as $ P^{\dagger} G P = -G $.
For a more satisfactory argument, we regard the momentum operator
as the generator of translation (see Eq. (\ref{eqn:2.14})).
We define the translation operator
\begin{equation}
	T(\lambda) := \exp( -\k \lambda G ).
	\label{eqn:2.38}
\end{equation}
It is geometrically obvious that translation followed by parity is equivalent
to parity followed by translation in the {\it opposite} direction.
Hence we should postulate that
\begin{equation}
	P \, T(\lambda) = T(-\lambda) \, P,
	\label{eqn:2.39}
\end{equation}
which implies
\begin{equation}
	P G = -G P ,
	\label{eqn:2.40}
\end{equation}
thus
\begin{equation}
	\mbox{(iv)} \qquad P^{\dagger} G P = -G.
	\label{eqn:2.41}
\end{equation}
Owing to the items (iii) and (iv), parity preserves the fundamental algebra
$ [\: G, W \:] = W, \; [\: G, W^{\dagger} \:] = -W^{\dagger} $ ;
phrased in a more refined language,
parity is an automorphism of the algebra $ {\cal A} $.
Apparently, the Hamiltonian (\ref{eqn:2.30}) is invariant under parity :
\begin{equation}
	P^{\dagger} H P = H.
	\label{eqn:2.42}
\end{equation}
\par
As already stated, an operator must have an operand;
namely, an operator should act on the Hilbert space
to be established as a quantum-mechanical operator.
Hence our next task is to represent the parity operator $ P $
on $ {\cal H}_{\alpha} $.
Applying $ G P $ on $ |n + \alpha \ket $
and using the property (\ref{eqn:2.40}), we obtain
\begin{eqnarray}
	G P |n + \alpha \ket
	& = &
	- P G |n + \alpha \ket
	\nonumber \\
	& = &
	- ( n + \alpha ) P |n + \alpha \ket .
	\label{eqn:2.43}
\end{eqnarray}
Hence
\begin{equation}
	P |n + \alpha \ket = c_{n + \alpha} |- n - \alpha \ket
	\label{eqn:2.44}
\end{equation}
with a coefficient $ c_{n + \alpha} $.
Notice that the right-hand side (RHS) is {\it not}
an element of $ {\cal H}_{\alpha} $
if $ \alpha $ is neither an integer $ m $ nor a half-integer $ m+\frac 12 $.
Only when $ \alpha = m $ or $ m + \frac 12 $,
$ P $ is well-defined on $ {\cal H}_{\alpha} $.
Otherwise, $ P $ must be represented on
$ {\cal H}_{\alpha} \oplus {\cal H}_{-\alpha} $.
This peculiar situation is similar to the {\it spontaneous symmetry breaking}
(SSB) in quantum field theory, in which
a symmetry of Hamiltonian is not a symmetry of a ground state
and the generator of the broken symmetry must be an ill-defined operator.
It has been believed that SSB can occur only
in a system with infinite degrees of freedom.
However, here we are treating a system with just one degree of freedom!
It is rather speculative
that the ill-defined parity in QM on $ S^1 $ has a connection with
SSB in quantum field theory.
\par
Let us return to the first problem, that is the representation of $ P $.
Suppose that $ \alpha = m $ or $ m + \frac 12 $.
Since $ P $ is unitary, $ |c_{n + \alpha}|=1 $.
Moreover $ P^2 = 1 $, thus
\begin{equation}
	c_{n + \alpha} \, c_{- n - \alpha} = 1 .
	\label{eqn:2.45}
\end{equation}
On the other hand, application of $ W P $ on $ |n + \alpha \ket $ gives
\begin{eqnarray}
	W P |n + \alpha \ket
	& = &
	c_{n + \alpha} W |- n - \alpha \ket
	\nonumber \\
	& = &
	c_{n + \alpha} |- n + 1 - \alpha \ket
	\nonumber \\
	& = &
	c_{n + \alpha} |-(n - 1 + \alpha) \ket
	\nonumber \\
	& = &
	c_{n + \alpha} \, c^*_{n - 1 + \alpha} \, P | n - 1 + \alpha \ket
	\nonumber \\
	& = &
	c_{n + \alpha} \, c^*_{n - 1 + \alpha} \,
	P W^{\dagger} | n + \alpha \ket ,
	\label{eqn:2.46}
\end{eqnarray}
which is equal to $ P W^{\dagger} |n + \alpha \ket $
by virtue of (\ref{eqn:2.37}). Accordingly, we obtain
\begin{equation}
	c_{n + \alpha} \, c^*_{n - 1 + \alpha} = 1,
	\label{eqn:2.47}
\end{equation}
which implies $ c_{n + \alpha} = c_{n - 1 + \alpha} $,
that is, all of $ c_{n + \alpha} $'s are equal to a constant $ c_{\alpha} $.
Therefore equation (\ref{eqn:2.45}) is reduced to
$ c_{\alpha}^2 = 1 $, thus $ c_{\alpha} = \pm 1 $.
Choice of $ c_{\alpha}=-1 $ is unitary equivalent to choice of
$ c_{\alpha}=1 $.
We choose the latter :
\begin{equation}
	P |n + \alpha \ket = |- n - \alpha \ket.
	\label{eqn:2.48}
\end{equation}
\par
For completeness, we shall consider a case such that
$ \alpha \neq m, m + \frac 12 $.
In this case, $ P $ operates on
$ {\cal H}_{\alpha} \oplus {\cal H}_{-\alpha} $.
The same consideration leading to (\ref{eqn:2.47})
leaves two arbitrary constants
$ c_{n + \alpha}=c_{\alpha}, \, c_{n - \alpha}=c_{- \alpha} $.
Since they are related by (\ref{eqn:2.45}), $ c_{- \alpha} = c_{\alpha}^* $.
It is readily seen that any choice of $ c_{\alpha} $ defines
a unitary equivalent representation.
If we write $ c_{\alpha} = \e^{\sk \delta} $, then
\begin{eqnarray}
	&&
	P | n + \alpha \ket
	=
	\e^{\sk \delta} | -n - \alpha \ket,
	\nonumber
	\\
	&&
	P | -n - \alpha \ket
	=
	\e^{-\sk \delta} | n + \alpha \ket.
	\label{eqn:2.49}
\end{eqnarray}
The unitary transformation
\begin{eqnarray}
	&&
	| n + \alpha \ket
	\to
	\e^{\frac 12 \sk \delta}
	| n + \alpha \ket ,
	\nonumber
	\\
	&&
	| -n - \alpha \ket
	\to
	\e^{- \frac 12 \sk \delta}
	| -n - \alpha \ket
	\label{eqn:2.50}
\end{eqnarray}
reduces (\ref{eqn:2.49}) to a simple representation :
\begin{eqnarray}
	&&
	P | n + \alpha \ket
	= | -n - \alpha \ket,
	\nonumber
	\\
	&&
	P | -n - \alpha \ket
	= | n + \alpha \ket.
	\label{eqn:2.51}
\end{eqnarray}
\par
How does a position eigenstate (\ref{eqn:2.9}) transform under parity?
First, suppose that $ \alpha = m $ or $ m + \frac 12 $.
A straightforward calculation is carried out as
\begin{eqnarray}
	P |\theta \ket
	& = &
	\kappa(\theta) \sum_n \e^{-\sk n \theta} |-n - \alpha \ket
	\nonumber
	\\
	& = &
	\kappa(\theta) \sum_n
	\e^{-\sk (-n - 2\alpha) (-\theta) + 2 \sk \alpha \theta}
	|(-n - 2\alpha)+\alpha \ket
	\nonumber
	\\
	& = &
	\e^{2 \sk \alpha \theta}
	\kappa(\theta) \sum_n
	\e^{-\sk n (-\theta) }
	|n + \alpha \ket
	\nonumber
	\\
	& = &
	\e^{2 \sk \alpha \theta}
	\kappa(\theta) \, \kappa^*(-\theta)
	| -\theta \ket .
	\label{eqn:2.52}
\end{eqnarray}
Specifically, if we set
\begin{equation}
	\kappa(\theta) =
	\left\{
	\begin{array}{ll}
		\e^{-\sk \alpha \theta}
		&
		\qquad \mbox{for} \: \alpha = m ,
		\\
		\e^{-\sk (\alpha - \frac 12) \theta}
		&
		\qquad \mbox{for} \: \alpha = m + \frac 12
	\end{array}
	\right.
	\label{eqn:2.53}
\end{equation}
(remember the arbitrariness of $ \kappa(\theta) $), we obtain
\begin{eqnarray}
	&&
	P |\theta \ket =
	\left\{
	\begin{array}{ll}
		|-\theta \ket
		&
		\qquad \mbox{for} \: \alpha = m ,
		\\
		\e^{\sk \theta} |-\theta \ket
		&
		\qquad \mbox{for} \: \alpha = m + \frac 12 .
	\end{array}
	\right.
	\label{eqn:2.54}
\end{eqnarray}
Notice that we {\it cannot} wipe away the phase factor $ \e^{\sk \theta} $
of the second equation by gauge transformation (\ref{eqn:2.22}).
\par
Secondly, let us consider a case of $ \alpha \neq m, m + \frac 12 $.
In this case we put an index $ \alpha $ on a position eigenstate
in order to indicate the Hilbert space containing the state :
\begin{equation}
	| \theta \ket_{\alpha} =
	\kappa_{\alpha} (\theta)
	\sum_{n=-\infty}^{\infty} \e^{-\sk n \theta} | n + \alpha \ket .
	\label{eqn:2.55}
\end{equation}
Using (\ref{eqn:2.51}), a calculation is carried out as
\begin{eqnarray}
	P |\theta \ket_{\alpha}
	& = &
	\kappa_{\alpha} (\theta) \sum_n \e^{-\sk n \theta} |-n - \alpha \ket
	\nonumber
	\\
	& = &
	\kappa(\theta)_{\alpha} \sum_n
	\e^{-\sk (-n) (-\theta) }	|(-n) - \alpha \ket
	\nonumber
	\\
	& = &
	\kappa_{\alpha} (\theta) \,
	\kappa^*_{-\alpha} (-\theta)
	| -\theta \ket_{-\alpha}.
	\label{eqn:2.56}
\end{eqnarray}
Setting $ \kappa_{\alpha} (\theta) = \kappa_{-\alpha} (-\theta) $, we obtain
\begin{equation}
	P | \theta \ket_{\alpha} = | -\theta \ket_{-\alpha} .
	\label{eqn:2.57}
\end{equation}
\par
We close this section by writing the parity transformation of a wave function.
It immediately follows (\ref{eqn:2.54}) that
\begin{equation}
	\psi(\theta) \to
	P \psi(\theta) :=
	\bra \theta | P | \psi \ket =
	\left\{
	\begin{array}{ll}
		\psi(-\theta)
		&
		\qquad \mbox{for} \: \alpha = m ,
		\\
		\e^{-\sk \theta} \psi(-\theta)
		&
		\qquad \mbox{for} \: \alpha = m + \frac 12 .
	\end{array}
	\right.
	\label{eqn:2.58}
\end{equation}
When $ \alpha \neq m, m + \frac 12 $, we array two components
of a wave function for
$ | \psi \ket \in {\cal H}_{\alpha} \oplus {\cal H}_{-\alpha} $ as
\begin{equation}
	\psi(\theta) =
	\left(
	\begin{array}{l}
		\psi_+ (\theta)
		\\
		\psi_- (\theta)
	\end{array}
	\right)
	:=
	\left(
	\begin{array}{l}
		{}_{\alpha}\bra \theta | \psi \ket
		\\
		{}_{-\alpha}\bra \theta | \psi \ket
	\end{array}
	\right)
	\to
	P \psi(\theta) =
	\left(
	\begin{array}{l}
		\psi_- (-\theta)
		\\
		\psi_+ (-\theta)
	\end{array}
	\right)
	.
	\label{eqn:2.59}
\end{equation}
%
%%%%%%%%%%%%%%%%%%%%%%%%%%%%%%%%%%%%%%%%%%%%%%%%%% sect 3 %%%%%%%%%%%%
%
\section{Uncertainty Relation}
\subsection{Usual argument}
In this section we review the derivation of the uncertainty relation
that can be found in the standard textbooks \cite{Sakurai}.
By doing this we will clarify our point of view.
\par
Assume that $ A $ and $ B $ are Hermitian operators.
Let $ | \psi \ket $ be a normalized state vector.
The expectation value of $ A $ is defined by
\begin{equation}
	\bra A \ket := \bra \psi | A | \psi \ket .
	\label{eqn:3.1}
\end{equation}
We introduce an operator
\begin{equation}
	\Delta A := A - \bra A \ket .
	\label{eqn:3.2}
\end{equation}
The variance of $ A $ is defined by
\begin{equation}
	\bra \Delta A^2 \ket =
	\bra A^2 \ket - \bra A \ket^2 .
	\label{eqn:3.3}
\end{equation}
The uncertainty relation of $ A $ and $ B $ is stated as
\begin{equation}
	\bra \Delta A^2 \ket \: \bra \Delta B^2 \ket
	\geq \frac 14 \, | \bra \, [\, A , B \,] \, \ket |^2 .
	\label{eqn:3.4}
\end{equation}
We will prove this inequality and examine when the sign of equality holds.
\par
For any vectors $ | \alpha \ket, \: | \beta \ket $ and any complex number
$ \lambda $, the positive-definiteness of the norm implies
\begin{equation}
	( \bra \alpha | + \lambda^* \bra \beta | )
	\cdot
	( | \alpha \ket + \lambda   | \beta \ket ) \geq 0 .
	\label{eqn:3.5}
\end{equation}
The sign of equality holds if and only if
\begin{equation}
	| \alpha \ket + \lambda | \beta \ket = 0 .
	\label{eqn:3.6}
\end{equation}
If that is the case, $ \lambda $ is given by
\begin{equation}
	\lambda =
	- \, \frac{ \bra \beta | \alpha \ket }{ \bra \beta | \beta \ket }.
	\label{eqn:3.7}
\end{equation}
The inequality (\ref{eqn:3.5}) must be true
even when we substitute (\ref{eqn:3.7}). Substitution gives
\begin{equation}
	\bra \alpha | \alpha \ket \: \bra \beta | \beta \ket
	\geq
	| \bra \alpha | \beta \ket |^2 ,
	\label{eqn:3.8}
\end{equation}
which is called the Schwarz inequality.
\par
Using (\ref{eqn:3.8}) with
\begin{equation}
	| \alpha \ket = \Delta A | \psi \ket , \qquad
	| \beta  \ket = \Delta B | \psi \ket ,
	\label{eqn:3.9}
\end{equation}
we obtain
\begin{equation}
	\bra \Delta A^2 \ket \: \bra \Delta B^2 \ket
	\geq
	| \bra \Delta A \, \Delta B \ket |^2 .
	\label{eqn:3.10}
\end{equation}
For the RHS, we note that
\begin{eqnarray}
	\Delta A \, \Delta B
	&=&
	\frac 12 \, \{  \Delta A , \Delta B  \} +
	\frac 12 \, [\, \Delta A , \Delta B \,]
	\nonumber
	\\
	&=&
	\frac 12 \, \{ \Delta A , \Delta B  \} +
	\frac 12 \, [\,       A ,        B \,]  ,
	\label{eqn:3.11}
\end{eqnarray}
where \{ , \} denotes the anti-commutator.
Obviously,
$ \{ \, \Delta A , \Delta B \, \} $ and $ [ \, A ,B \, ] $
are Hermitian and anti-Hermitian respectively.
Because the expectation value of an Hermitian operator is a real number and
the one of an anti-Hermitian operator is a purely imaginary number,
the RHS of (\ref{eqn:3.10}) is rewritten as
\begin{equation}
	| \bra \Delta A \, \Delta B \ket |^2
	=
	\frac 14 \, | \bra \, \{ \Delta A , \Delta B  \} \, \ket |^2 +
	\frac 14 \, | \bra \, [\,       A ,        B \,] \, \ket |^2  .
	\label{eqn:3.12}
\end{equation}
Omission of the first term of the RHS of (\ref{eqn:3.12}) makes the inequality
(\ref{eqn:3.10}) stronger and gives (\ref{eqn:3.4}).
\par
Next, we shall examine the condition for the sign of equality of
(\ref{eqn:3.4}) to be realized.
First, equation (\ref{eqn:3.6}) must be satisfied.
Secondly, the omitted term $ \bra \, \{ \Delta A , \Delta B \} \, \ket $
must vanish. Eqs. (\ref{eqn:3.6}) and (\ref{eqn:3.9}) imply
\begin{equation}
	\bra \psi | \{ \Delta A , \Delta B \} | \psi \ket
	=
	- ( \lambda + \lambda^* ) \,
	\bra \psi | \Delta B^2 | \psi \ket .
	\label{eqn:3.13}
\end{equation}
Thus we conclude that the sign of equality of (\ref{eqn:3.4}) holds
if and only if
\begin{equation}
	( \Delta A + \lambda \Delta B ) | \psi \ket = 0
	\label{eqn:3.14}
\end{equation}
is satisfied with a purely imaginary number $ \lambda $.
If these conditions are satisfied,
we call $ | \psi \ket $ a minimum uncertainty state.
\subsection{Uncertainty relation on $ S^1 $}
Now we are in a place to study our main subject.
We fix the representation space $ {\cal H}_{\alpha} $.
Applying the Schwarz inequality (\ref{eqn:3.8}) to
\begin{equation}
	| \alpha \ket = \Delta G | \psi \ket , \qquad
	| \beta  \ket = \Delta W | \psi \ket ,
	\label{eqn:3.15}
\end{equation}
we obtain
\begin{equation}
	\bra \Delta G^{\: 2} \ket \: \bra \Delta W^\dagger \Delta W \ket
	\geq
	| \bra \Delta G \, \Delta W \ket |^2 .
	\label{eqn:3.16}
\end{equation}
Notice that $ W $ is not Hermitian.
Thus we cannot naively apply (\ref{eqn:3.4}) to $ A=G $ and $ B=W $.
\par
The sign of equality of (\ref{eqn:3.16}) holds
if (\ref{eqn:3.14}) is satisfied, that is
\begin{equation}
	\Bigl( G - \bra G \ket \Bigr) | \psi \ket
	= - \lambda \Bigl( W - \bra W \ket \Bigr) | \psi \ket .
	\label{eqn:3.17}
\end{equation}
In terms of wave function, this equation is expressed as
\begin{equation}
	\Bigl(-\k \frac{\partial}{\partial \theta} + \alpha
	- \bra G \ket \Bigr)
	\psi (\theta)
	=
	- \lambda \Bigl( \e^{\sk \theta} - \bra W \ket \Bigr)
	\psi (\theta),
	\label{eqn:3.18}
\end{equation}
where we have used (\ref{eqn:2.17}) and (\ref{eqn:2.18}),
setting $ \kappa(\theta) \equiv 1 $.
The solution of (\ref{eqn:3.18}) is
\begin{equation}
	\psi (\theta)
	= N \exp
	\left[
	- \lambda \e^{\sk \theta}
	+ \k\Bigl( - \alpha + \bra G \ket + \lambda \bra W \ket \Bigr) \theta
	\right] .
	\label{eqn:3.19}
\end{equation}
To ensure that $ \psi(\theta + 2 \pi) = \psi(\theta) $,
\begin{equation}
	\nu := - \alpha + \bra G \ket + \lambda \bra W \ket
	\label{eqn:3.20}
\end{equation}
must be an integer.
Moreover, it is required that $ \bra G \ket $ is a real number.
Therefore, if we write $ \bra W \ket = \rho \, \e^{\sk \varphi} $,
there must be a real number $ \beta $ such that
\begin{equation}
	\lambda = - \beta \, \e^{-\sk \varphi} ,
	\label{eqn:3.21}
\end{equation}
where we put the minus sign in front of $ \beta $ for later convenience.
Accordingly, (\ref{eqn:3.19}) is rewritten as
\begin{equation}
	\psi (\theta)
	= N \exp
	\left[
	\beta \, \e^{\sk ( \theta - \varphi ) }
	+ \k \nu \theta
	\right] .
	\label{eqn:3.22}
\end{equation}
We call it a minimum uncertainty wave packet on $ S^1 $.
It is apparent that the probability density
defined by the above wave function :
\begin{equation}
	| \psi (\theta) |^2
	= | N |^2 \exp \left[ 2 \beta \cos ( \theta - \varphi ) \right]
	\label{eqn:3.23}
\end{equation}
has its peak at $ \theta = \varphi $.
$ N $ is determined by the normalization condition :
\begin{eqnarray}
	1
	& = &
	\bra \psi | \psi \ket
	\nonumber \\
	& = &
	| N |^2
	\int_0^{2 \pi} \frac{\di \theta}{2 \pi} \,
	\e^{2 \beta \cos (\theta - \varphi)}
	\nonumber \\
	& = &
	| N |^2 \:
	I_0 (2 \beta),
	\label{eqn:3.24}
\end{eqnarray}
where $ I_n(z) $ is the $ n $-th modified Bessel function.
Explanation of the modified Bessel functions is given in the appendix.
The parameter $ \beta $ is related to $ \rho $ by
\begin{eqnarray}
	\rho \, \e^{\sk \varphi}
	& = &
	\bra \psi | W | \psi \ket
	\nonumber \\
	& = &
	| N |^2
	\int_0^{2 \pi} \frac{\di \theta}{2 \pi} \,
	\e^{2 \beta \cos (\theta - \varphi)} \, \e^{\sk \theta}
	\nonumber \\
	& = &
	| N |^2 \:
	I_1 (2 \beta) \, \e^{\sk \varphi},
	\label{eqn:3.25}
\end{eqnarray}
which implies
\begin{equation}
	\rho = \frac{ I_1 (2 \beta) }{ I_0 (2 \beta) }.
	\label{eqn:3.26}
\end{equation}
The consistency condition $ \bra G \ket = \bra \psi | G | \psi \ket $
is also satisfied by the above choice of parameters.
Here we rearrange the parameters for definiteness :
\begin{eqnarray}
	\bra W \ket
	& = &
	\rho \, \e^{\sk \varphi},
	\label{eqn:3.27}
	\\
	\bra G \ket
	& = &
	\nu + \alpha + \beta \rho .
	\label{eqn:3.28}
\end{eqnarray}
$ \beta $ and $ \rho $ are related by the equation (\ref{eqn:3.26}),
which can be solved for $ \beta $;
the inverse function $ \beta = \beta (\rho) $ is a single-valued function for
$ 0 \le \rho \le 1 $.
Thus the remaining arbitrary parameters are
$ \rho \, ( 0 \le \rho \le 1 ), \: \nu \, (=0, \pm 1, \pm 2, \cdots ) $ and
$ \varphi $.
\par
We proceed to evaluate the variances of $ W $ and $ G $.
For the position, we have
\begin{eqnarray}
	\bra \Delta W^\dagger \Delta W \ket
	& = &
	\bra \: ( W^\dagger - \bra W \ket^* ) ( W - \bra W \ket ) \: \ket
	\nonumber \\
	& = &
	1 - \bra W \ket^* \bra W \ket
	\nonumber \\
	& = &
	1 - \rho^2 .
	\label{eqn:3.29}
\end{eqnarray}
Notice that this equation holds for any state $ | \psi \ket $.
Of course, the variance $ \bra \Delta W^\dagger \Delta W \ket $ is bounded.
For the momentum, using (\ref{eqn:3.17}) and (\ref{eqn:3.21}), we obtain
\begin{eqnarray}
	\bra \Delta G^{\: 2} \ket
	& = &
	\beta^2 \bra \Delta W^\dagger \Delta W \ket
	\nonumber \\
	& = &
	\beta^2 ( 1 - \rho^2 ).
	\label{eqn:3.30}
\end{eqnarray}
Therefore the product of two variances is
\begin{equation}
	\bra \Delta G^{\: 2} \ket \: \bra \Delta W^\dagger \Delta W \ket
	=
	\beta^2 ( 1 - \rho^2 )^2 .
	\label{eqn:3.31}
\end{equation}
Next we consider two extreme cases ; (i) $ \rho \to 0 $, (ii) $ \rho \to 1 $.
\par
(i) It is obvious that the limit $ \rho \to 0 $
makes the minimum uncertainty state a momentum eigenstate.
For a small $ z $, it is known that
\begin{eqnarray}
	I_0 (z)
	& \approx &
	1 + \frac{ z^2 }{ 4 } + \cdots ,
	\label{eqn:3.32}
	\\
	I_1 (z)
	& \approx &
	\frac{z}{2} + \frac{ z^3 }{16} + \cdots
	\label{eqn:3.33}
\end{eqnarray}
(see the appendix, Eq. (\ref{eqn:a.1})).
Hence (\ref{eqn:3.26}) implies that $ \rho \approx \beta $ as $ \beta \to 0 $.
Therefore we conclude that
$ \bra \Delta W^\dagger \Delta W \ket \to 1 $ and
$ \bra \Delta G^{\: 2} \ket \to 0 $ for the limit $ \rho \to 0 $.
\par
(ii) It is expected that the limit $ \rho \to 1 $
makes the minimum uncertainty state a position eigenstate.
For a large $ z $, it is known that
\begin{eqnarray}
	I_0 (z)
	& \approx &
	\frac 1{\sqrt{2 \pi z}}
	\: \e^z
	\left[
	1 + \frac 1{8 z} + \cdots
	\right],
	\label{eqn:3.34}
	\\
	I_1 (z)
	& \approx &
	\frac 1{\sqrt{2 \pi z}}
	\: \e^z
	\left[
	1 - \frac 3{8 z} + \cdots
	\right]
	\label{eqn:3.35}
\end{eqnarray}
(also see the appendix, Eqs. (\ref{eqn:a.8}) and (\ref{eqn:a.9})).
Using them, it is readily seen that
\begin{equation}
	\frac{ I_1(z) }{ I_0(z) }
	\approx
	1 - \frac{1}{ 2z }.
	\label{eqn:3.36}
\end{equation}
Therefore (\ref{eqn:3.26}) behaves asymptotically for large $ \beta $ as
\begin{equation}
	\rho
	=
	\frac{ I_1 (2 \beta) }{ I_0 (2 \beta) }
	\approx
	1 - \frac 1{ 4 \beta },
	\label{eqn:3.37}
\end{equation}
which implies that $ \rho \to 1 $ as $ \beta \to \infty $. Thus we have
\begin{equation}
	1 - \rho^2
	=
	( 1 + \rho ) ( 1 - \rho )
	\approx
	2 \cdot \frac 1{ 4 \beta }
	=
	\frac 1{ 2 \beta }.
	\label{eqn:3.38}
\end{equation}
Consequently we deduce that as $ \beta \to \infty $
\begin{eqnarray}
	\bra \Delta W^\dagger \Delta W \ket
	& \approx &
	\frac 1{ 2 \beta },
	\label{eqn:3.39}
	\\
	\bra \Delta G^{\: 2} \ket
	& \approx &
	\frac {\beta}{2},
	\label{eqn:3.40}
\end{eqnarray}
and that
\begin{equation}
	\bra \Delta W^\dagger \Delta W \ket \:
	\bra \Delta G^{\: 2} \ket
	\approx
	\frac 14.
	\label{eqn:3.41}
\end{equation}
\par
It should be noticed that the parameter $ \alpha $ is irrelevant
to the above consideration.
It only shifts $ \bra G \ket $ as indicated by (\ref{eqn:3.28}).
\subsection{Large radius limit}
Up to here we have assumed that the radius of circle is fixed.
Now suppose that the radius $ r $ can be varied.
If $ r $ grows up,
an arbitrary part of the arc gets close to a straight segment.
It is naively expected that QM on $ S^1 $ will be reduced to the ordinary
QM on a line when $ r $ increases to infinity.
It is also expected that the ordinary uncertainty relation
$ \Delta x \Delta p \ge \frac 12 \hbar $
will be recovered.\footnote{In this paper we have put $ \hbar=1 $.}
Here we consider this issue;
the large radius limit of the minimum uncertainty state.
\par
Observing Eq. (\ref{eqn:3.24}), we choose the normalization constant $ N $ as
\begin{equation}
	N = \frac 1{ \sqrt{ I_0(2\beta) }} \, \e^{-\sk \nu \varphi}.
	\label{eqn:3.42}
\end{equation}
We rewrite (\ref{eqn:3.22}) here :
\begin{equation}
	\psi (\theta)
	=
	\frac 1{ \sqrt{ I_0(2\beta) }} \,
	\exp
	\left[
	\beta \, \e^{\sk ( \theta - \varphi ) }
	+ \k ( \bra G \ket - \alpha - \beta \rho ) ( \theta - \varphi )
	\right] .
	\label{eqn:3.43}
\end{equation}
We define variables $ x, \, \bra x \ket $ and $ \bra p \ket $ by
\begin{eqnarray}
	x & := & r \theta ,
	\nonumber
	\\
	\bra x \ket & := & r \varphi ,
	\label{eqn:3.44}
	\\
	\bra p \ket & := & \frac 1r \, \bra G \ket .
	\nonumber
\end{eqnarray}
We would like to change the argument of the wave function from
$ \theta $ to $ x $.
Introducing the normalization convention
\begin{equation}
	| \psi (\theta) |^2
	\: \frac{ \di \theta }{ 2 \pi }
	=
	| \Psi (x) |^2
	\, \di x ,
	\label{eqn:3.45}
\end{equation}
we define a rescaled wave function as
\begin{eqnarray}
	\Psi(x)
	&:= &
	\frac 1{ \sqrt{ 2 \pi r}} \: \psi(\theta)
	\nonumber
	\\
	& = &
	\frac 1{ \sqrt{ 2 \pi r \, I_0(2\beta) }} \,
	\exp
	\left[
	\beta \, \cos \Bigl( \frac{ x - \bra x \ket }{ r } \Bigr)
	+ \k \beta \, \sin \Bigl( \frac{ x - \bra x \ket }{ r } \Bigr)
	\right.
	\nonumber
	\\
	& & \qquad \qquad \qquad
	\left.
	+ \k ( r \bra p \ket - \alpha - \beta \rho )
	\Bigl( \frac{ x - \bra x \ket }{ r } \Bigr)
	\right] .
	\label{eqn:3.46}
\end{eqnarray}
Noticing the asymptotic behavior
\begin{equation}
	I_n (z)
	\approx
	\frac 1{\sqrt{2 \pi z}} \: \e^z
	\qquad ( z \gg n, 1 ),
	\label{eqn:3.47}
\end{equation}
and taking the limit $ \rho \to 1 \, ( \beta \to \infty ), \, r \to \infty $,
(\ref{eqn:3.46}) is reduced to
\begin{eqnarray}
	\Psi(x)
	& \approx &
	\left(
	\frac { \beta }{ \pi r^2 }
	\right)^{\frac 14}
	\e^{ - \beta } \,
	\exp
	\left[
	\beta
	- \frac 12 \beta \Bigl( \frac{ x - \bra x \ket }{ r } \Bigr)^2
	+ \k \beta \Bigl( \frac{ x - \bra x \ket }{ r } \Bigr)
	\right.
	\nonumber
	\\
	& & \qquad \qquad \qquad
	\left.
	+ \k ( r \bra p \ket - \alpha - \beta \rho )
	\Bigl( \frac{ x - \bra x \ket }{ r } \Bigr)
	\right]
	\nonumber
	\\
	& = &
	\left(
	\frac { \beta }{ \pi r^2 }
	\right)^{\frac 14}
	\exp
	\left[
	- \frac { \beta }{ 2 r^2 } \, ( x - \bra x \ket )^2
	+ \k \bra p \ket ( x - \bra x \ket )
	\right.
	\nonumber
	\\
	& & \qquad \qquad \qquad
	\left.
	+ \k \beta ( 1 - \rho ) \Bigl( \frac{ x - \bra x \ket }{ r } \Bigr)
	- \k \alpha \Bigl( \frac{ x - \bra x \ket }{ r } \Bigr)
	\right] .
	\label{eqn:3.48}
\end{eqnarray}
The last term of the last line vanishes as $ r \to \infty $.
Because of (\ref{eqn:3.37}): $ \beta ( 1 - \rho ) \approx \frac 14 $,
the term just in front of the last one also vanishes as $ r \to \infty $.
If we take the limit, keeping the relation
\begin{equation}
	\frac{ \beta }{ r^2 } = \frac 1{ 2 d^2 } = \mbox{const} ,
	\label{eqn:3.49}
\end{equation}
we arrive at
\begin{equation}
	\Psi(x)
	=
	\left( \frac 1{ 2 \pi d^2 } \right)^{\frac 14}
	\exp
	\left[
	- \frac 1{ 4 d^2 } ( x - \bra x \ket )^2
	+ \k \bra p \ket    ( x - \bra x \ket )
	\right] ,
	\label{eqn:3.50}
\end{equation}
which is nothing but a Gaussian wave packet.
We can substitute arbitrary real numbers for
$ \bra x \ket $ and $ \bra p \ket $.
It is well known that the uncertainty product for it is
$ \bra \Delta x^2 \ket \bra \Delta p^2 \ket = \frac 14 $,
which is consistent with (\ref{eqn:3.41}).
Hence we conclude that the minimum uncertainty wave packet on $ S^1 $
results in a Gaussian wave packet on a line
when the large radius limit is performed.
%
%%%%%%%%%%%%%%%%%%%%%%%%%%%%%%%%%%%%%%%%%%%%%%%%%% sect 4 %%%%%%%%%%%%
%
\section{Discussions}
\subsection{Summary}
Here we summarize what we have studied in this paper.
We reviewed QM on $ S^1 $ formulated by Ohnuki and Kitakado.
They began with definition of the fundamental algebra for it
and constructed irreducible representations of the algebra.
They have shown that there are many inequivalent representations,
which are parametrized by $ \alpha \: (0 \le \alpha < 1)$.
They also showed that a gauge field that leads to various QM is built in.
The gauge field is interpreted as the vector potential for magnetic flux
$ 2 \pi \alpha $
penetrating $ S^1 $ embedded in higher dimension.
It is already known that the gauge field on $ S^1 $ does not influence
a classical motion but does a quantum-mechanical amplitude.
\par
We found that the energy spectrum of a free particle depends on $ \alpha $.
The spectrum exhibits degeneracies and a symmetry for the special value of
$ \alpha $, that is an integer or a half-integer.
This symmetry is identified with parity.
We defined the parity operator and constructed its representations.
We found the situation where it becomes ill-defined
and this situation reminded us of SSB in quantum field theory.
\par
The uncertainty relation between position and momentum on $ S^1 $
was considered.
A minimum uncertainty state is constructed.
Both of a certain momentum limit and a certain position limit are examined.
Finally, we showed that
when the radius of $ S^1 $ increases infinitely,
the minimum uncertainty wave packet on $ S^1 $
results in a Gaussian wave packet on a line
and the usual uncertainty relation is recovered.
\par
In the rest of this paper,
we would like to state some remarks including some speculations.
\subsection{Global aspect of quantum theory}
I would like to emphasize that operators are global objects.
To illustrate my point of view I take an instructive example
from Shiga's book \cite{Shiga}.
\par
Let us consider an eigenvalue problem of the differential operator
\begin{equation}
	- \frac{ \di^2 u }{ \di x^2} = \lambda u
	\label{eqn:4.1}
\end{equation}
with a boundary condition $ u(0) = u(1) = 0 $.
The solution is
\begin{eqnarray}
	&&
	\lambda_n = n^2 \, \pi^2 ,
	\nonumber
	\\
	&&
	u_n(x) = a_n \sin n \pi x \qquad (n=1,2,3,\cdots).
	\label{eqn:4.2}
\end{eqnarray}
We can rewrite the differential equation (\ref{eqn:4.1})
to an integral equation as follows.
Integrating (\ref{eqn:4.1}) twice, we obtain
\begin{equation}
	u(x) =
	-\lambda
	\int_0^x \di s \,
	\int_0^s \di t \, u(t)
	+ c_1 x + c_2 .
	\label{eqn:4.3}
\end{equation}
$ u(0)=0 $ implies $ c_2=0 $.
Integrating by part, we obtain
\begin{eqnarray}
	\int_0^x \di s \,
	\int_0^s \di t \, u(t)
	&=&
	\int_0^x \di s
	\left(
		1 \cdot	\int_0^s \di t \, u(t)
	\right)
	\nonumber
	\\
	&=&
	x \int_0^x \di t \, u(t)
	- \int_0^x \di s \, s \, u(s)
	\nonumber
	\\
	&=&
	\int_0^x \di t \, (x - t) u(t).
	\label{eqn:4.4}
\end{eqnarray}
Thus (\ref{eqn:4.3}) is
\begin{equation}
	u(x) =
	- \lambda \int_0^x \di t \, (x - t) u(t) + c_1 x.
	\label{eqn:4.5}
\end{equation}
Since $ u(1)=0 $,
\begin{equation}
	c_1 =
	\lambda \int_0^1 \di t \, (1 - t) u(t) .
	\label{eqn:4.6}
\end{equation}
Substitution of this into (\ref{eqn:4.5}) gives
\begin{eqnarray}
	u(x)
	&=&
	- \lambda \int_0^x \di t \, (x - t) u(t)
	+ \lambda \int_0^1 \di t \, x (1 - t) u(t)
	\nonumber
	\\
	&=&
	- \lambda \int_0^x \di t \, (x - t) u(t)
	+ \lambda \int_0^x \di t \, x (1 - t) u(t)
	+ \lambda \int_x^1 \di t \, x (1 - t) u(t)
	\nonumber
	\\
	&=&
	\lambda \left(
		  \int_0^x \di t \, t (1 - x) u(t)
		+ \int_x^1 \di t \, x (1 - t) u(t)
	\right).
	\label{eqn:4.7}
\end{eqnarray}
If we define
\begin{equation}
	Q(x,t) :=
	\left\{
	\begin{array}{ll}
		t (1-x) & \qquad \mbox{when  } 0 \le t \le x, \\
		x (1-t) & \qquad \mbox{when  } x \le t \le 1, \\
	\end{array}
	\right.
	\label{eqn:4.8}
\end{equation}
equation (\ref{eqn:4.7}) is rewritten as
\begin{equation}
	\hat{Q} u(x) :=
	\int_0^1 \di t \, Q(x,t) u(t) = \frac 1{\lambda} u(x),
	\label{eqn:4.9}
\end{equation}
which is an eigenvalue problem of the integral operator $ \hat{Q} $.
$ Q(x,t) $ is called a Green function
for the boundary value problem (\ref{eqn:4.1}).
Notice that the boundary condition $ u(0) = u(1) = 0 $ is automatically ensured
by the definition of $ Q(x,t) $.
Inversely, one can easily verify that the solution of (\ref{eqn:4.9})
satisfies (\ref{eqn:4.1}).
\par
Therefore we have seen that the differential equation (\ref{eqn:4.1})
with the boundary condition
is equivalent to the integral equation (\ref{eqn:4.9}).
The above example tells us that operators have a global nature;
here by the term ``a global nature'' I mean that
the operator $ \hat{Q} $ makes sense by integration
over the whole interval [ 0,1 ]
and that the boundary condition is already involved in $ \hat{Q} $.
Cutting the interval [ 0,1 ] into pieces deprives $ \hat{Q} $ of sense.
This global nature also obstructs formulation of QM on manifolds
as is seen in what follows.
\par
To bring the contrast
we would like to turn our attention to classical mechanics.
Classical mechanics describes dynamics of a system
in terms of a differential equation,\footnote{Although there are also
discrete dynamical systems, which are described by difference equations,
we do not consider them here.}
which is expressed by some coordinate.
The Lagrangian formalism permits us to use a wide class of coordinates,
namely, generalized coordinates on a configuration space.
Furthermore, the Hamiltonian formalism permits usage
of a still wider class of coordinates,
namely, canonical coordinates on a phase space.
More generally, differential geometry offers the most flexible framework
to classical mechanics;
the whole of states of a dynamical system forms a manifold,
an equation of motion is a vector field on the manifold,
dynamics is an integral curve of the vector field
and observables are functions on the manifold.
All objects---states, dynamics and observables---are
{\it defined locally and patched globally.}
\par
Now we return to QM.
Both of state vectors and observables have the global nature;
a state vector grasps a system as a whole, not separated pieces of it.
For an observable, we have already seen that an operator has the global nature.
As another example to show this nature, let us consider the usual canonical
commutation relation of self-adjoint operators $ x_j, p_j $ :
\begin{equation}
	[\: x_j , x_k \:] = [\: p_j , p_k \:] = 0,
	\qquad
	[\: x_j , p_k \:] = \k \delta_{jk}.
	\label{eqn:4.10}
\end{equation}
It immediately follows self-adjoitness and the algebra that
$ x_j $ and $ p_j $ have continuous spectra ranging over $ (-\infty,\infty) $.
Hence QM on an Euclidean space is enforced on us.
The operator formalism of QM lacks flexibility which
differential geometry has;
freedom to choose coordinates and paste them onto a manifold.
I guess that self-adjointness rather than the algebra
is responsible for the global nature.
\par
Path integral also exhibits a global nature;
when we studied path integral for QM on $ S^1 $,
sum over paths is performed weighting a path winding $ n $ times
with the phase factor $ \exp( -\k \alpha 2 \pi n ) $.
It attracted our attention that
topological concepts such as
the winding number and the weight factor for homotopy classes have appeared.
\par
Through the above investigations,
we have observed a profound character of QM, that is a global aspect of QM.
Classical mechanics does not have a counterpart.
\par
When we expand our discussion into quantum field theory,
we encounter a global aspect again, for example, the Wess-Zumino-Witten term.
Furthermore, as a field-theoretical version of QM on a manifold,
we already have a nonlinear sigma model,
which treats {\it a manifold-valued field}
to describe dynamics of Nambu-Goldstone bosons at low energy \cite{Bando}.
Here questions arise;
can we quantize a manifold-valued field in the operator formalism?
Is it valid that we naively use the canonical commutation relation
of field-variables as the fundamental algebra?
Can we construct a representation space?
Does construction of the Fock space still hold good?
According to Kamefuchi-O'Raifeartaigh-Salam theorem
\cite{Bando}, \cite{Kamefuchi},
the {\it S}-matrix is independent of the choice of field variables
{\it as far as the canonical quantization and the perturbative method
are valid.}
Unfortunately, a global aspect cannot be seen by perturbative approach.
We expect to construct quantum theory for a manifold-valued field seriously.
This problem, that is quantization of a manifold-valued field,
attracts an academic interest.
Furthermore, we expect that we will find another peculiar character of nature
by solving it.
Moreover, we remark that quantization of
{\it a manifold-valued field on a manifold} seems to be an ambitious attempt.
\par
Before closing this section, we note that a differential geometric approach
to QM is already proposed by several authors \cite{Bayen}, \cite{Batalin}.
Furthermore, we have already found another way
to construct QM on more general manifolds.
We will publish a detailed discussion in a subsequent paper \cite{Tani2}.
\subsection{Observables in quantum mechanics on manifolds}
What quantities are observables in QM on $ S^1 $?
In particular, what quantities can be used
to indicate the position of a particle?
The angle coordinate $ \theta $ cannot be observable,
because it is not single-valued.
We have used the unitary operator $ W $ to specify a point on $ S^1 $
by eigenvalues of $ W $ (see Eq.(\ref{eqn:2.10})).
However its eigenvalues range over complex numbers $ \{ \e^{\sk \theta} \} $.
If one feels uncomfortable against use of complex numbers
for observable quantity, we offer a substitution for $ W $;
we introduce the cosine and sine operators as
\begin{equation}
	C := \frac 12 ( W + W^{\dagger} ),
	\qquad
	S := \frac 1{2 \, \k} \, ( W - W^{\dagger} ),
	\label{eqn:4.11}
\end{equation}
which satisfy the following commutation relations
\begin{eqnarray}
	&&
	[ \: G , C \: ] = \k \, S,
	\label{eqn:4.12}
	\\
	&&
	[ \: G , S \: ] = - \k \, C,
	\label{eqn:4.13}
	\\
	&&
	[ \: C , S \: ] = 0,
	\label{eqn:4.14}
\end{eqnarray}
and obeys a constraint
\begin{equation}
	C^2 + S^2 = 1.
	\label{eqn:4.15}
\end{equation}
The spectra of $ C $ and $ S $ range over an interval $ [ -1,1 ] $.
Inversely, if we begin with self-adjoint operators $ G, C $ and $ S $
which satisfy the above algebra from (\ref{eqn:4.12}) to (\ref{eqn:4.15}),
we can reconstruct QM on $ S^1 $.
Using the cosine and sine operators, Carruthers and Nieto \cite{Carruthers}
have discussed the uncertainty relation on $ S^1 $
However, they did not recognize existence of an infinite number of
inequivalent QM.
\par
The above consideration suggests that
when we attempt to formulate QM on a general manifold,
we encounter the same problem; {\it do position operators exist?}
Here position operators are defined as a set of operators whose eigenvalues
have one-to-one correspondence to points on a manifold.
A general manifold cannot be covered with a single coordinate-patch.
Even a sphere needs at least two coordinate-patches;
for example, $ S^1 $ needs cosine and sine.
As emphasized in the previous section, operators are global objects.
The need of plural coordinate-patches conflicts with
the global nature of operators.
Probably, for QM on a manifold,
we should give up use of position operators to specify a point.
\par
If we discard position operators, what remains observable?
We expect that probability density remains observable quantity indicating
a position of a particle.
Spectrum of Hamiltonian, transition amplitudes will also remain observable.
Yet we have not considered a condition to decide
what are physically observable.
\par
Concerning the uncertainty of position on $ S^1 $,
we have used the expectation value $ \bra \Delta W^\dagger \Delta W \ket $
to measure it.
On the other hand,
Judge \cite{Judge} has introduced another quantity to measure it.
Here we quote his argument for comparison.
Let $ \psi(\theta) $ be a wave function on $ S^1 $ :
$ \psi(\theta + \pi) = \psi(\theta) $.
$ | \psi(\theta) |^2 $ gives a probability distribution on $ S^1 $.
He introduced
\begin{equation}
	\Delta \theta^2 :=
	\min_{-\pi \le \gamma \le \pi}
	\left\{
	\int_{-\pi}^{\pi} \theta^2 \,| \psi( \gamma + \theta) |^2 \: \di \theta
	\right\}
	\label{eqn:4.16}
\end{equation}
and defined $ \Delta \theta := \sqrt{ \Delta \theta^2 } $.
This quantity $ \Delta \theta $ has the following properties:
\begin{enumerate}
	\renewcommand{\labelenumi}{(\roman{enumi})}
	\item
		$ \Delta \theta $ is independent
		of the choice of origin of angle coordinate.
	\item
		$ \Delta \theta = 0 $
		for a sharp ($ \delta $-function) distribution.
	\item
		$ \Delta \theta = \pi / \sqrt{ 3 } $
		for a uniform distribution.
\end{enumerate}
Furthermore, he defined the angular momentum operator
\begin{equation}
	L := - \k \hbar \, \frac{\partial}{\partial \theta}
	\label{eqn:4.17}
\end{equation}
and he defined the expectation value and the uncertainty of it as usual:
\begin{eqnarray}
	\bra L \ket
	& := &
	\int_{-\pi}^{\pi} \psi^*(\theta) L \, \psi(\theta) \: \di \theta ,
	\label{eqn:4.18}
	\\
	\Delta L^2
	& := &
	\int_{-\pi}^{\pi} \psi^*(\theta) \,
	[ L - \bra L \ket ]^2            \,
	\psi(\theta) \: \di \theta.
	\label{eqn:4.19}
\end{eqnarray}
He conjectured that the uncertainty relation between them should be
\begin{equation}
	\Delta L \cdot
	\frac{ \Delta \theta }
	{ 1 - (3/\pi^2) \Delta \theta^2 }
	\ge
	\frac 12 \hbar ,
	\label{eqn:4.20}
\end{equation}
and Bouten et al. \cite{Bouten} proved his conjecture.
It should be noticed that Judge did not use the angle variable $ \theta $
as the position operator.
He used the probability distribution as a basic quantity instead.
\subsection{Remaining problems}
Finally, we propose three problems.
\par
{\it 1. Dynamical systems :}
In this paper, we have taken only a free particle
as a concrete system on $ S^1 $.
We expect to study other dynamical systems, for example,
{\it a quantum pendulum}, whose Hamiltonian is
\begin{equation}
	H = \frac 12 G^{\: 2} - \frac g2 ( W + W^{\dagger} ).
	\label{eqn:4.21}
\end{equation}
As a natural extension of it, {\it a spherical pendulum} offers another
exercise to QM on $ S^D $.
\par
{\it 2. Statistics of identical particles :}
When there are plural particles on a manifold, what statistics do they obey?
Can we find an answer in the context of QM?
Or otherwise, should we resort to quantum field theory
to consider this subject?
Laidlaw and DeWitt \cite{Laidlaw} have considered it
from a viewpoint of path integral.
\par
{\it 3. Gravity in quantum mechanics :}
We know that the concept of gauge fields neatly fits QM.
The role of vector potential is recognized essentially in the context of QM,
for example, in phenomena such as the Aharonov-Bohm effect
and the Berry phase \cite{Berry}, \cite{Shapere}.
Ohnuki and Kitakado also showed that introduction of non-Abelian gauge field
is inevitable for QM on $ S^D $.
Historically, the concept of gauge symmetry leads us
to deep understanding of interactions found in nature.
The gauge principle plays a role of guide in physics.
\par
On the other hand, we have the equivalence principle
as a guide to understanding the gravitational interaction.
However, the gravitational field does not have
a comfortable room in QM.\footnote{Here we do not refer to the difficulty
of so-called quantum gravity.
We do not attempt to quantize the gravitational field itself;
we are concerned to introduce neatly it into QM of a particle.}
Greenberger \cite{Greenberger1}, \cite{Greenberger2}
and Staudenmann et al. \cite{Staudenmann}
have discussed the role of the equivalence principle in QM.
\par
We shall compare the manner of gravity to appear in QM
with that in classical mechanics.
Here we quote Sakurai's argument \cite{Sakurai}.
The classical equation of motion for a purely falling body is
\begin{equation}
	m \, \frac{\di^2 \bm{x}}{\di t^2}
	= - m \, \nabla \Phi_{\mbox{\scriptsize grav}},
	\label{eqn:4.22}
\end{equation}
where $ \Phi_{\mbox{\scriptsize grav}} $ is the gravitational potential.
Since the inertial mass is equal to the gravitational mass,
the coefficient $ m $, that is the mass, drops out.
Therefore the individual character of the body is irrelevant to its motion.
The peculiarity of gravity is universality.
Because of its universality, gravity in classical mechanics can be reduced to
a pure geometry of the space-time.
\par
The situation is different in QM.
The Schr{\"o}dinger equation for the falling body is
\begin{equation}
	\k \hbar \frac{\partial \psi}{\partial t} =
	\left[
	- \frac{\hbar^2}{2 m} \nabla^2
	+ m \, \nabla \Phi_{\mbox{\scriptsize grav}}
	\right]
	\psi.
	\label{eqn:4.23}
\end{equation}
The mass no longer cancels; instead it appears in the combination $ \hbar/m $.
Thus in a problem where $ \hbar $ appears, $ m $ is also expected to appear.
To put it briefly, quantization annoys universality of gravity.
\par
It is expected to find a way to bring the gravitational field into QM as neatly
as gauge fields are brought into play.
We have proposed one way \cite{Tanimura},
which introduces the gravitational field
in the commutation relation between position and velocity.
%
%%%%%%%%%%%%%%%%%%%% acknowledgments %%%%%%%%%%%%% sect 5 %%%%%%%%%%%%
%
\section*{Acknowledgments}
I am grateful to Prof. S. Kitakado and Prof. Y. Ohnuki
for suggestive discussions.
I would like to thank Prof. S. Kitakado and Prof. A.I. Sanda
for careful reading of my manuscript.
I would like to thank Dr. Shibata for his help on drawing graphs.
I am deeply indebted to Dr. Tsujimaru for his encouragement.
%
%%%%%%%%%%%%%%%%%%%% appendix %%%%%%%%%%%%%%%%%%%% sect 6 %%%%%%%%%%%%
%
\section*{Appendix : modified Bessel functions}
\setcounter{equation}{0}
\renewcommand{\theequation}{A.\arabic{equation}}
Here we summarize useful formulas on the Bessel functions \cite{Dictionary}.
The Bessel functions are defined by
\begin{eqnarray}
	J_n (z)
	&:=&
	\sum_{m=0}^{\infty} \frac{ (-1)^m }{ m! \, (n+m)! } \,
	\left( \frac z2 \right)^{n+2m}
	\qquad
	(n=0,1,2,\cdots) ,
	\nonumber
	\\
	J_{-n} (z)
	&:=&
	(-1)^n \, J_n (z).
	\label{eqn:a.1}
\end{eqnarray}
They have the integral representation :
\begin{equation}
	J_n (z)
	= \int_0^{2 \pi} \frac{\di \theta}{2 \pi} \, \exp
	\Bigl[
	-\k z \cos \theta + \k n ( \theta + \frac{\pi}2 )
	\Bigr]
	\qquad
	(n=0,\pm 1,\pm 2,\cdots) ,
	\label{eqn:a.2}
\end{equation}
which is called Hansen-Bessel's formula.
The modified Bessel function is defined as
\begin{equation}
	I_n (z)
	:=
	\e^{-\sk n \frac{\pi}2 } J_n ( \k z)
	\qquad
	(n=0,\pm 1,\pm 2,\cdots) .
	\label{eqn:a.3}
\end{equation}
Hence it is represented by series
\begin{equation}
	I_n (z)
	=
	\sum_{m=0}^{\infty} \frac{1}{ m! \, (n+m)! } \,
	\left( \frac z2 \right)^{n+2m}
	\qquad
	(n=0, 1, 2, \cdots) ,
	\label{eqn:a.3a}
\end{equation}
and also represented by integration
\begin{equation}
	I_n (z)
	= \int_0^{2 \pi} \frac{\di \theta}{2 \pi} \, \exp
	\Bigl[
	z \cos \theta + \k n \theta
	\Bigr]
	\qquad
	(n=0,\pm 1,\pm 2,\cdots) .
	\label{eqn:a.4}
\end{equation}
When $ |z| \gg |n| $ and $ |z| \gg 1 $, the asymptotic behavior of $ J_n(z) $
is given by Hankel's asymptotic series :
\begin{eqnarray}
	J_n (z)
	& = &
	\sqrt{\frac 2{\pi z}}
	\cos \Bigr( z - \frac{n \pi}2 - \frac{\pi}4 \Bigr)
	\left[
	\sum_{m=0}^{M-1} (-1)^m \, \frac{(n,2m)}{(2z)^{2m}}
	+ O( |z|^{-2M} )
	\right]
	\nonumber \\
	& &
	- \sqrt{\frac 2{\pi z}}
	\sin \Bigr( z - \frac{n \pi}2 - \frac{\pi}4 \Bigr)
	\left[
	\sum_{m=0}^{M-1} (-1)^m \, \frac{(n,2m+1)}{(2z)^{2m+1}}
	+ O( |z|^{-2M-1} )
	\right]
	\nonumber \\
	& &
	\qquad \qquad
	( -\pi < \arg z < \pi ) ,
	\label{eqn:a.5}
\end{eqnarray}
where the symbol $ (n,m) $ is defined as
\begin{eqnarray}
	(n,m)
	&:=&
	\frac{ [4n^2-1^2] \, [4n^2-3^2] \cdots
	[4n^2-(2m-1)^2] }{ 2^{2m} \: m!}
	\quad
	(m=1,2,3,\cdots) ,
	\nonumber \\
	(n,0)
	&:=& 1 .
	\label{eqn:a.6}
\end{eqnarray}
Eqs. (\ref{eqn:a.3}) and (\ref{eqn:a.5}) gives an asymptotic series
for $ I_n(z) $ :
\begin{equation}
	I_n (z) =
	\frac 1{\sqrt{2 \pi z}}
	\: \e^z
	\left[
	\sum_{m=0}^{2M-1} (-1)^m \, \frac{(n,m)}{(2z)^m}
	+ O( |z|^{-2M} )
	\right]
	\quad
	( - \frac 12 \pi < \arg z < \frac 12 \pi ) .
	\label{eqn:a.7}
\end{equation}
In particular, the concrete forms for $ n=0,1 $ are
\begin{eqnarray}
	I_0 (z)
	& \approx &
	\frac 1{\sqrt{2 \pi z}}
	\: \e^z
	\left[
	1 + \frac 1{8 z} + \frac 9{128 z^2} + \cdots
	\right],
	\label{eqn:a.8}
	\\
	I_1 (z)
	& \approx &
	\frac 1{\sqrt{2 \pi z}}
	\: \e^z
	\left[
	1 - \frac 3{8 z} - \frac {15}{128 z^2} + \cdots
	\right].
	\label{eqn:a.9}
\end{eqnarray}
%
%%%%%%%%%%%%%%%%%%%% bibliography %%%%%%%%%%%%%%%% sect 7 %%%%%%%%%%%%
%
%\newpage

%
\end{document}